\documentclass{osa-article}

\journal{oe}

\articletype{Research Article}

\begin{document}
	
	\title{Experimental self-testing for photonic graph states}
	
	\author{Jia-Min Xu,\authormark{1} Qing Zhou,\authormark{2} Yu-Xiang Yang,\authormark{3,4} Zi-Mo Cheng, \authormark{3,4} Xin-Yu Xu, \authormark{1} Zhi-Cheng Ren,\authormark{3,4} Xi-Lin Wang\authormark{3,4,*}and Hui-Tian Wang,\authormark{3,4,5}}
	
	\address{\authormark{1}Department of Modern Physics, University of Science and Technology of China, Hefei 230026, China}
	
	\address{\authormark{2}Hefei National Laboratory for Physical Sciences at Microscale, University of Science and Technology of China,  Hefei 230026,  China}
	
	\address{\authormark{3}National Laboratory of Solid State Microstructures, School of Physics, Nanjing University, Nanjing 210093, China}
	
	\address{\authormark{4}Collaborative Innovation Center of Advanced Microstructures, Nanjing 210093, China}
		
		\email{\authormark{5}e-mail: htwang@nju.edu.cn} 
		\email{\authormark{*}Corresponding author: xilinwang@nju.edu.cn} 
	
	

\begin{abstract}
\noindent
Graph states---one of the most representative families of multipartite entangled states, are important resources for multiparty quantum communication, quantum error correction, and quantum computation. Device-independent certification of highly entangled graph states plays a prominent role in the quantum information processing tasks. Here we have experimentally demonstrated device-independent certification for multipartite graph states, by adopting the robust self-testing scheme based on scalable Bell inequalities. Specifically, the prepared multi-qubit Greenberger-Horne-Zeilinger (GHZ) states and linear cluster states achieve a high degree of Bell violation, which are beyond the nontrivial bounds of the robust self-testing scheme. Furthermore, our work paves the way to the device-independent certification of complex multipartite quantum states. 
\end{abstract}

\section{Introduction}
In the field of quantum physics, multipartite entanglement is certainly viewed as a crucial resource~\cite{PhysRev.47.777,pan_rmp_multiphoton,wang_prl_2016,zhong_prl_2018,Takesue:09,Zhu:20}. In many practical information processing tasks, multipartite entanglement provides more advantages than bipartite one. Multipartite entanglement not only facilitates the information processing, but also has lots of applications in quantum simulation~\cite{Lloyd1073} and quantum error correction~\cite{365700,chiaverini2004realization}. In addition, multipartite entanglement can be used to reveal nonlocality and entanglement property by violating the particular Bell inequalities~\cite{pan2000experimental,RevModPhys.86.419,PhysRevLett.108.200401}. Therefore, it is of great significance to develop the device-independent tools for certifying the genuine multipartite entanglement.  

The Bell nonlocality test is one of the most important and interesting device-independent tools~\cite{brunner2014bell}. In this scenario, assuming that a pair of unknown quantum states are distributed to two parties (Alice and Bob) and each party performs local measurements, the unique information they can obtain is some settings $(x, y)$ and outcomes $(a, b)$ of measurement apparatus, which can be adopted to estimate the correlation of observed statistics $P(a,b|x,y)$. By observing the violation of a certain Bell inequality with these statistics, one can test whether the unknown state shared between Alice and Bob is entangled.  The Bell nonlocality test, as a device-independent method, has been widely applied in various quantum information tasks. More powerful testing methods for describing an unknow quantum states have been developed.  

The method of inferring more detailed properties of a quantum experiment in a black-box scenario is referred to as ``self-testing''~\cite{Supic2020selftestingof,PhysRevLett.117.070402,PhysRevA.98.062307,PhysRevA.87.050102,PhysRevA.98.052115,jebarathinam2019maximal,vsupic2019device,goswami2018one,sarkar2019self,bharti2019robust,li2020self,li2018self,vsupic2016self,wu2016device,coopmans2019robust,natarajan2016robust}. Mayer and Yao~\cite{10.5555/2011827.2011830} first proposed such a device-independent method for certifying any type of quantum system. The self-testing method can be used to certify not only particular quantum states, such as all pure bipartite entangled states~\cite{coladangelo2017all}, three-qubit W states~\cite{wu2014robust}, graph states~\cite{mckague2011self} and generalized GHZ (Greenberger-Horne-Zeilinger) states~\cite{vsupic2018self}, but also various sets of measurements such as Bell state measurement~\cite{PhysRevLett.121.250506,PhysRevLett.121.180505}, unsharp measurement~\cite{miklin2020semi} and non-projection measurement~\cite{PhysRevLett.117.260401,Tavakolieaaw6664}. Self-testing method has also some extended applications to other scenarios, for example, the certification of a quantum gate or unitary transformation in a device-independent manner~\cite{magniez2006self, Micuda:17}. Recently, it has also been proven that entangled subspaces are self-testable~\cite{PhysRevLett.125.260507}.

From the theoretical point of view, there are lots of results in self-testing. However, due to the imperfect experimental systems, one cannot achieve the ideal self-testing results, i.e., the most self-testing methods are still only theoretical recipe. To realize the self-testing task in laboratory, many robust self-testing protocols have been developed to tolerate certain noise, in particular, some of them have also been successfully demonstrated in optical experiment~\cite{Tavakolieaaw6664,PhysRevLett.121.240402,PhysRevLett.122.090402,wang2018multidimensional,zhang2019experimental,PRXQuantum2020346}. However, the experimental realization of self-testing for multipartite entanglement states has not been demonstrated yet. Recently, Baccari \textit{et al}.~\cite{Baccari_2020} have developed a new method to derive scalable Bell inequalities for graph states, based on the knowledge of stabilizers. Meanwhile, these new scalable Bell inequalities can be used to self-test various graph states. Compared with the previous constructions of Bell inequalities, the proposed scalable Bell inequalities show that the number of expected values they require to measure changes linearly with $N$ ($N$ is number of qubits). Fewer correlators make them amenable for experimental realization. Based on the numerical results in Ref.~\cite{Baccari_2020}, the scalable Bell inequalities are suitable for nontrivial self-testing graph states under certain noise.

\begin{figure}[!htp]
\centering
\includegraphics[width=.85\linewidth]{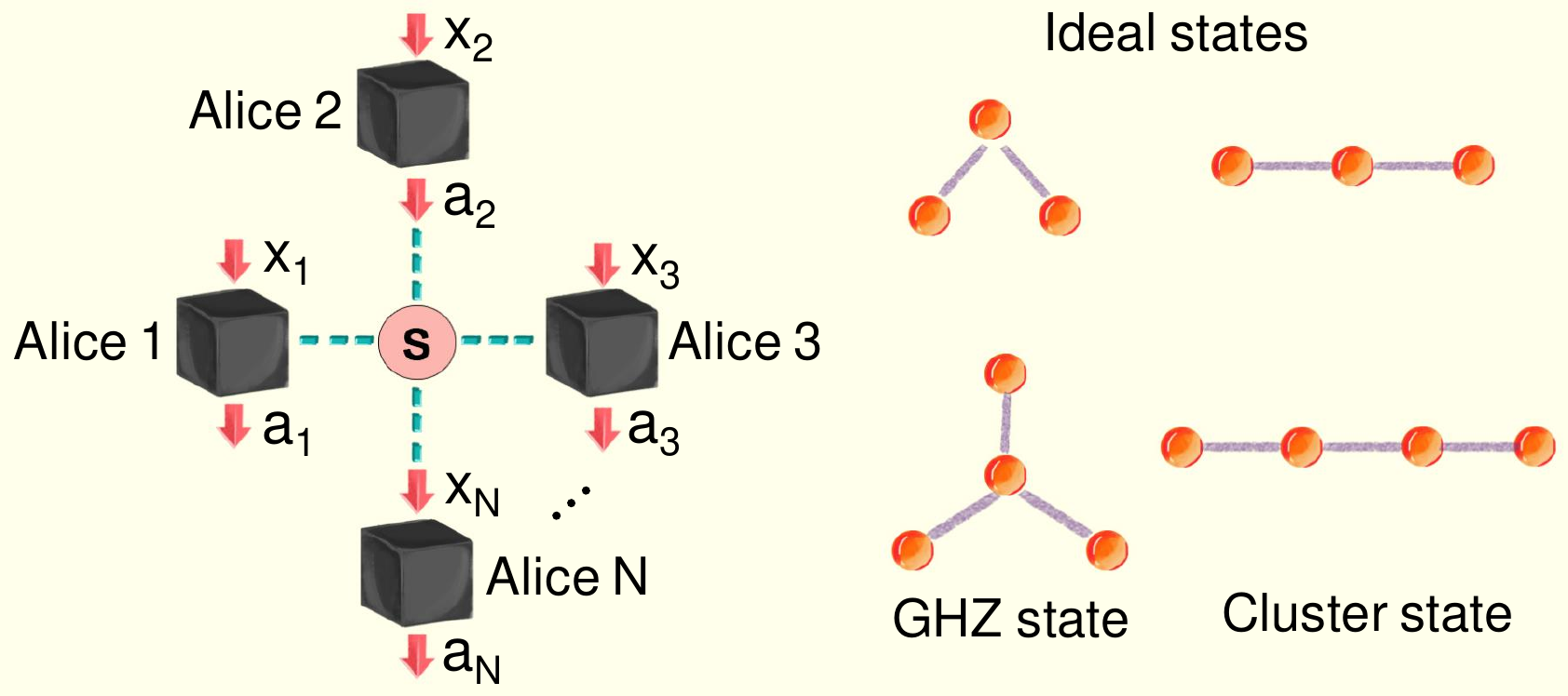} 
\caption{Self-testing for multipartite graph states. (a) The schematic illustration of self-testing: an unknown multipartite quantum state in central node is distributed to Alice 1, Alice 2, $\dots$, Alice $N$. $\{x_1\sim x_N\}\in\{0,1\}$ and $\{a_1\sim a_N\}\in\{0,1\}$ are measurements inputs (settings) and outputs respectively. One can estimate the correlations $P(a_1, a_2, \dots, a_N|x_1, x_2, \dots, x_N)$ by these settings and outputs. If the obtained correlations can achieve the violation of certain Bell inequality, these correlations can be used to self-test target state. Here, we use the scalable Bell inequalities in Ref.~\cite{Baccari_2020} to self-test the multipartite graph states. (b) The ideal states are self-tested: three- and four-partite GHZ states and cluster states.}
\label{fig:scheme}
\end{figure}

Here, based on the scalable Bell inequalities, we experimentally investigate robust self-testing for two important types of multipartite graph states---GHZ states and linear cluster states, with four-photon entanglement sources. We demonstrate the self-testing for multi-qubit GHZ and cluster states under experimental noise. The results indicate that these scalable Bell inequalities are nontrivial to robustly self-test various multipartite graph states.

\section{Theoretical model}
\subsection{General scalable Bell inequalities}
Firstly, let us recall the self-testing task. Assuming that there is an unknown source as shown in Fig.~\ref{fig:scheme}(a) distributed to $N$ parties. Each party can be seen as a black-box, and has no internal information about measurement devices. $N$ parties can acquire information of inputs (settings) and outputs, and then obtain some correlations. The goal of self-testing is to device-independently reveal the structure of unknown states distributed by the unknown source from these correlations only. The method presented firstly in Ref.~\cite{Baccari_2020} utilizes the certain violation of multipartite Bell inequality to self-test the graph states. The explicit form of multipartite Bell inequalities is written as follows

\noindent
\begin{equation}
\begin{aligned}
I_G=  &n_{\mathrm{max}}\left \langle (A_0^{(1)} + A_1^{(1)})\prod\limits_{i\in n(1)} A_1^{(i)} \right \rangle   +\sum\limits_{i\in n(1)} \left \langle (A_0^{(1)}-A_1^{(1)})A_0^{(i)}\prod\limits_{j\in n(i) \backslash\{1\}}A_1^{(j)} \right \rangle \\
& +\sum\limits_{i\notin n(1)\cup\{1\}}\left \langle A_0^{(i)}\prod\limits_{j\in n(i)}A_1^{(j)} \right \rangle \leq \beta_G^C ,
\end{aligned}
\end{equation}

\noindent where $n(i)$ is the set of neighborhood for the $i$th vertex in the graph $G$, $n_{\mathrm{max}}\equiv max_{i}|n(i)|$, and $|n(i)|$ denotes the number of the set $n(i)$. $A_{0}^{(i)}$ and $A_{1}^{(i)}$ are the possible measurements performed on the $i$th party. The classical bound and the maximum quantum violation in Eq.~(1) are $\beta_G^C=n_{\mathrm{max}}+N-1$ and $\beta_G^Q=(2\sqrt2-1)n_{\mathrm{max}}+N-1$, respectively \cite{Baccari_2020}.

\subsection{GHZ states}
First of all, for $N$-qubit GHZ state, the stabilizing operators denote $G_1=X_1 \cdots X_N$ and $G_i=Z_1Z_i$ with $i=2,\dots,N$. Such a GHZ state is local-unitary equivalent to the star graph with $\otimes_{i=2}^N H$. By performing unitary transformation, one can obtain the following Bell inequality~\cite{Baccari_2020}.

\begin{equation}
\begin{aligned}
I_{GHZ}^N=&(N-1)\left(\left \langle A_0^{(1)}A_0^{(2)} \cdots A_0^{(N)} \right \rangle+ \left \langle A_1^{(1)}A_0^{(2)} \cdots A_0^{(N)} \right \rangle \right) \\
&+\sum\limits_{i=2}^N\left(\left \langle A_0^{(1)}A_1^{(i)}\right \rangle- \left \langle A_1^{(1)}A_1^{(i)} \right \rangle \right)\leq 2(N-1). 
\end{aligned}
\end{equation}

For a $N$-qubit GHZ state, we have $n_{\mathrm{max}}= N-1$, thus its maximum violation should be $\beta_{GHZ}^Q=2\sqrt{2}(N-1)$. To self-test the quality of GHZ state in laboratory, the measurement settings in the Bell inequality are $A_{0}^{(1)}=(X+Z)/\sqrt{2}$, $A_{1}^{(1)}=(X-Z)/\sqrt{2}$ and $A_{0}^{(i)}=X, A_{1}^{(i)}=Z$ with $i \geq 2$. If the maximum violation can be obtained with these measurement settings, the unknown state is equivalent to a GHZ state, up to a local isometry. When considering noise for $N=3$ ($N=4$) as shown in Fig.~\ref{fig:scheme}(b), from the numerical results in Ref.~\cite{Baccari_2020}, a violation beyond 4.828 (7.464) implies that the fidelity of the prepared state with respect to an ideal GHZ state exceeds 0.5.

Following the recipe, the Bell inequality of three-qubit GHZ state can be deduced from Eq. (2) as

\begin{equation}
\begin{aligned}
I_{GHZ}^3 = & 2\left( \left \langle A_0^{(1)}A_0^{(2)}A_0^{(3)}\right \rangle + \left \langle A_1^{(1)}A_0^{(2)}A_0^{(3)}\right \rangle \right) + \left \langle A_0^{(1)}A_1^{(2)}\right \rangle \\
& - \left \langle A_1^{(1)}A_1^{(2)}\right \rangle + \left \langle A_0^{(1)}A_1^{(3)}\right \rangle - \left \langle A_1^{(1)}A_1^{(3)}\right \rangle \leq 4
\end{aligned}
\end{equation}

Following the \textbf{Fact 3} in Ref.~\cite{Baccari_2020}, the observables are equivalent up to $A_0^{(1)}=[X + Z]/\sqrt{2}$, $A_1^{(1)}=[X - Z]/\sqrt{2}$, and $A_{0/1}^{(i)}=X/Z$ for $i=2, \dots, N$. So the concrete form of Bell inequality are

\begin{equation}
\begin{aligned}
I_{GHZ}^3 = & 2\left(\left \langle \frac{X+Z}{\sqrt2}XX \right \rangle + \left \langle \frac{X-Z}{\sqrt2}XX \right \rangle\right)  +\left \langle \frac{X+Z}{\sqrt2}ZI \right \rangle  \\
 -&\left \langle \frac{X-Z}{\sqrt2}ZI \right \rangle  + \left \langle \frac{X+Z}{\sqrt2}IZ \right \rangle - \left \langle \frac{X-Z}{\sqrt2}IZ \right \rangle
\end{aligned}
\end{equation}

For the four-qubit GHZ state, we have

\begin{equation}
\begin{aligned}
I_{GHZ}^4 = & 3\left( \left \langle A_0^{(1)}A_0^{(2)} \cdots A_0^{(N)}\right \rangle + \left \langle A_1^{(1)}A_0^{(2)} \cdots A_0^{(N)}\right \rangle \right) \\
&+ \left \langle A_0^{(1)}A_1^{(2)}\right \rangle - \left \langle A_1^{(1)}A_1^{(2)}\right \rangle + \left \langle A_0^{(1)}A_1^{(3)}\right \rangle \\
&- \left \langle A_1^{(1)}A_1^{(3)}\right \rangle +  \left \langle A_0^{(1)}A_1^{(4)}\right \rangle - \left \langle A_1^{(1)}A_1^{(4)}\right \rangle\leq 6
\end{aligned}
\end{equation}

The corresponding measurement operator should be
\begin{equation}
\begin{aligned}
I_{GHZ}^4 = & 3\left(\left \langle \frac{X+Z}{\sqrt2}XXX \right \rangle + \left \langle \frac{X-Z}{\sqrt2}XXX \right \rangle\right) + \left \langle \frac{X+Z}{\sqrt2}ZII \right \rangle - \left \langle \frac{X-Z}{\sqrt2}ZII \right \rangle \\
&+ \left \langle \frac{X+Z}{\sqrt2}IZI \right \rangle - \left \langle \frac{X-Z}{\sqrt2}IZI \right \rangle + \left \langle \frac{X+Z}{\sqrt2}IIZ \right \rangle - \left \langle \frac{X-Z}{\sqrt2}IIZ \right \rangle
\end{aligned}
\end{equation}

\subsection{Ring states}
For a $N$-qubit ring state, its stabilizing operators read $G_i=Z_{i-1}X_{i}Z_{i+1}$ for $i=1, \dots, N$, where $Z_0 \equiv Z_N$ and $Z_{N+1}\equiv Z_1$. As each vertex in the ring graph has the same size neighbourhood, namely, $|n(i)|=n_{\mathrm{max}}=2$ for $i=1, \dots, N$, one can derive the following Bell inequality
\begin{equation}
\begin{aligned}
I_{\mathrm{ring}}^{N}=&2\left \langle A_1^{(N)}(A_0^{(1)}+A_1^{(1)})A_1^{(2)} \right \rangle + \left \langle (A_0^{(1)}-A_1^{(1)})A_0^{(2)}A_1^{(3)}  \right \rangle \\
 &+ \left \langle  A_1^{(N-1)} A_0^{(N)} (A_0^{(1)}-A_1^{(1)}) \right \rangle + \sum\limits_{i=3}^{N-1} \left \langle A_1^{(i-1)}A_0^{(i)}A_1^{(i+1)}   \right \rangle \leq  N+1  .
\end{aligned}
\end{equation}

The maximum quantum violation is $\beta^Q_{\mathrm{ring}}=N+4\sqrt{2}-3$ for the $N$-qubit ring state. In experiment, we focus on the three-photon and four-photon situations. Specifically, the three-photon ($N=3$) ring state can be transformed into a linear cluster state through the unitary operation $(\sqrt{Z}\otimes\sqrt{X}\otimes\sqrt{Z})^{\dag}$. The four-photon ($N=4$) ring state can be converted into a linear cluster state by a relabeling of qubits 2 and 3 and the local unitary transformation $H_1\otimes H_2\otimes H_3\otimes H_4$~\cite{walther2005experimental}. Thus we have equalities $\beta^C_{\mathrm{cluster}}=\beta^C_{\mathrm{ring}}$ and $\beta^Q_{\mathrm{cluster}}=\beta^Q_{\mathrm{ring}}$ when $N=3,4$, as shown in Fig.~\ref{fig:scheme}(b). The measurement settings are the same as the GHZ case. The violation beyond 4.940 (5.828) implies the fidelity of prepared state to be an ideal three-qubit (four-qubit) cluster state exceeds 0.5~\cite{Baccari_2020}.

The form of ring states explicit as
\begin{equation}
\begin{aligned}
|R_3\rangle = & \frac{1}{2}(|HH+\rangle+|HV-\rangle+|VH-\rangle-|VV+\rangle) \\
|R_4\rangle = & \frac{1}{2}(|H+H+\rangle+|H-V-\rangle+|V-H-\rangle+|V+V+\rangle)
\end{aligned}
\end{equation}

When $N=3$, one reads
\begin{equation} 
\begin{aligned}
I_{\mathrm{ring}}^{3} = & 2\left \langle A_1^{(3)}(A_0^{(1)}+A_1^{(1)})A_1^{(2)} \right \rangle + \left \langle (A_0^{(1)}-A_1^{(1)})A_0^{(2)}A_1^{(3)} \right \rangle \\
& + \left \langle  A_1^{(2)} A_0^{(3)}   (A_0^{(1)}-A_1^{(1)})    \right \rangle \leq 4
\end{aligned}
\end{equation} 

Substituting $A_{0}^{(1)} = [X + Z]/\sqrt{2}$, $A_{1}^{(1)} = [X - Z]/\sqrt{2}$, $A_{0}^{(2)} = A_{0}^{(3)} = X$, and $A_{1}^{(2)} = A_{1}^{(3)} = Z$ into the above expression, we have

\begin{equation} 
\begin{aligned}
I_{\mathrm{ring}}^3 = & 2\left(\left \langle \frac{X+Z}{\sqrt2}ZZ \right \rangle + \left \langle \frac{X-Z}{\sqrt2}ZZ \right \rangle\right) +\left \langle \frac{X+Z}{\sqrt2}XZ \right \rangle \\
&- \left \langle \frac{X-Z}{\sqrt2}XZ \right \rangle + \left \langle \frac{X+Z}{\sqrt2}ZX \right \rangle - \left \langle \frac{X-Z}{\sqrt2}ZX \right \rangle
\end{aligned}
\end{equation}

Since the Bell inequality is insensitive to local unitary. Here, to simplify the experimental setup, we prepare the linear cluster states $|C_3\rangle$ and $|C_4\rangle$ substituted for the ring states $|R_3\rangle$ and $|R_4\rangle$. For the three-qubit ring state $|R_3\rangle$ can be converted into $|C_3\rangle=\frac{1}{\sqrt2}(|+H+\rangle + |-V-\rangle)$ through the unitary operation $(\sqrt Z \sqrt X \sqrt Z)^\dagger$. The measurement operator corresponding to $|R_3\rangle$ turns to 

\begin{equation}
\begin{aligned}
I_{\mathrm{cluster}}^3 = & 2\left(\left \langle \frac{Y+Z}{\sqrt2}YZ \right \rangle + \left \langle \frac{Y-Z}{\sqrt2}YZ \right \rangle\right)  +\left \langle \frac{Y+Z}{\sqrt2}XZ \right \rangle \\
&- \left \langle \frac{Y-Z}{\sqrt2}XZ \right \rangle + \left \langle \frac{Y+Z}{\sqrt2}YY \right \rangle - \left \langle \frac{Y-Z}{\sqrt2}YY \right \rangle
\end{aligned} 
\end{equation}

The Bell-like inequality of four-qubit ring state indicates
\begin{equation}
\begin{aligned}
I_{\mathrm{ring}}^4 = & 2\left( \left \langle A_0^{(1)}A_1^{(2)}...A_1^{(4)}\right \rangle + \left \langle A_1^{(1)}A_1^{(2)}...A_1^{(4)}\right \rangle \right) \\
&+ \left \langle A_0^{(1)}A_0^{(2)}A_1^{(4)}\right \rangle - \left \langle A_1^{(1)}A_0^{(2)}A_1^{(4)}\right \rangle + \left \langle A_1^{(3)}A_0^{(4)}A_0^{(1)}\right \rangle \\
& - \left \langle A_1^{(3)}A_0^{(4)}A_1^{(1)}\right \rangle + \left \langle A_1^{(2)}A_0^{(3)}A_1^{(4)}\right \rangle  \leq 5
\end{aligned}
\end{equation}

which is consistent with 
\begin{equation}
\begin{aligned}
I_{\mathrm{ring}}^4 = & 2\left(\left \langle \frac{X+Z}{\sqrt2}ZIZ \right \rangle + \left \langle \frac{X-Z}{\sqrt2}ZIZ \right \rangle\right)+\left \langle \frac{X+Z}{\sqrt2}XZI \right \rangle \\
&- \left \langle \frac{X-Z}{\sqrt2}XZI \right \rangle  + \left \langle \frac{X+Z}{\sqrt2}IZX \right \rangle - \left \langle \frac{X-Z}{\sqrt2}IZX \right \rangle +  \left \langle IZXZ \right \rangle
\end{aligned}
\end{equation}

The four-qubit ring state $|R_4\rangle$ can transform into $|C_4\rangle=\frac{1}{2}(|HHHH\rangle+|HHVV\rangle+|VVHH\rangle-|VVVV\rangle)$ by relabeling the qubits 2 and 3 and the local unitary operation $H_1 \otimes H_2 \otimes H_3 \otimes H_4$~\cite{walther2005experimental}. We need to accordingly change the measurement operator in Eq. 14
\begin{equation}
\begin{aligned}
I_{\mathrm{cluster}}^4 = & 2\left(\left \langle \frac{X+Z}{\sqrt2}IXX \right \rangle - \left \langle \frac{X-Z}{\sqrt2}IXX \right \rangle\right) +\left \langle \frac{X+Z}{\sqrt2}XZI \right \rangle \\
&+ \left \langle \frac{X-Z}{\sqrt2}XZI \right \rangle   + \left \langle \frac{X+Z}{\sqrt2}XIZ \right \rangle+ \left \langle \frac{X-Z}{\sqrt2}XIZ \right \rangle + \left \langle IZXX \right \rangle
\end{aligned}
\end{equation}

\section{Experimental setup}

\begin{figure*}[!hbt]
\centering
\includegraphics[width=0.95\linewidth]{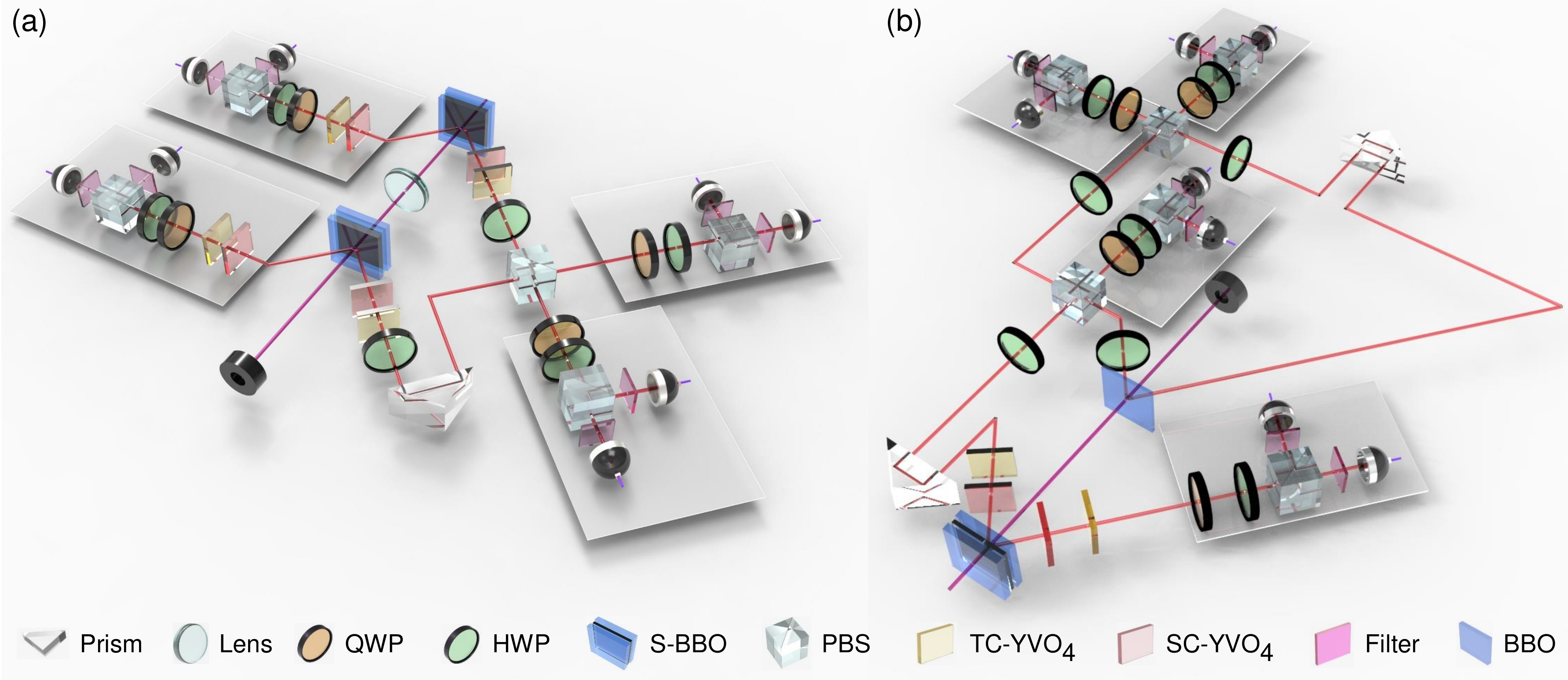}
\caption{Experimental setup for self-testing multipartite graph states. (a) Setup for generating two GHZ states and a three-photon cluster state. An ultrafast ultraviolet pulsed laser beam with a central wavelength of 390 nm, a duration of $\sim$140 fs and a repetition rate of 80 MHz, successively passes through the sandwich-like combination of a beamlike BBO crystal, a HWP and a beamlike BBO crystal to generate the entangled photon pairs $|\Psi^+\rangle=(|HV\rangle + |VH\rangle)/\sqrt{2}$.
(b) Setup for generating the four-photon cluster state. The second beamlike single BBO crystal is utilized to generate a correlated photon pair in the state of $|HV\rangle$, then the photons with polarization $H$ ($V$) pass through HWP at $22.5^\circ$ ($67.5^\circ$), resulting in the state of $|+\rangle = (|H\rangle + |V\rangle)/\sqrt{2}$. All photons are filtered with 3-nm bandwidth filter. QWP: quarter-wave plate, HWP: half-wave plate, BBO:  $\beta$-barium borate, S-BBO: sandwich structure of BBO+HWP+BBO, PBS: polarizing beam splitter, TC-YVO$_4$: birefringent YVO$_4$ crystal for temporal compensation (TC), and SC-YVO$_4$: birefringent YVO$_4$ crystal for spatial compensation (SC).}
\label{fig:fig2}
\end{figure*}

We prepare a variety of multipartite graph states, with entangled photons from spontaneous parametric down conversion (SPDC) by using the beamlike type-II S-BBO, which is a sandwich structure, i.e., a half-wave plate (HWP) at $45^\circ$ is sandwiched between two identical 2-mm-thick $\beta$-barium borate (BBO) crystals~\cite{wang_prl_2016}, as shown in Fig.~\ref{fig:fig2}(a). For a four-photon GHZ state, we firstly pump two S-BBOs to produce two photon pairs in the state $|\Psi^+\rangle=(|HV\rangle + |VH\rangle)/\sqrt{2}$ and insert a HWP in one photon path to convert the state into $|\Phi^+\rangle=(|HH\rangle + |VV\rangle)/\sqrt{2}$, where $|H\rangle$ and $|V\rangle$ denote the horizontal and vertical polarizations of single photons, respectively. Then, we use a polarizing beam splitter (PBS) to post-select the four-photon GHZ state $|G_4\rangle=(|HHHH\rangle+|VVVV\rangle)/\sqrt 2$. By projecting one of the four photons into $|+\rangle = (|H\rangle + |V\rangle)/\sqrt{2}$, we can conveniently generate a three-photon GHZ state $|G_3\rangle$. We can also prepare the three-qubit linear cluster state in a similar manner, as it is locally equivalent to the three-qubit GHZ state. We take three steps to generate the four-photon linear cluster state as shown in Fig.~\ref{fig:fig2}(b). We firstly prepare a photon pair in the Bell state $|\Phi^+\rangle$ and another photon pair in the state $|++\rangle$. Then we utilize a PBS to produce a three-photon GHZ state with the photon pair in the Bell state $|\Phi^+\rangle$ and one photon in the state $|+\rangle$ from the other photon pair. In the third step, we use another PBS to implement a parity check operation for the fourth photon in the state $|+\rangle$ and one photon from the three photons in the GHZ state. Based on the above operation, highly entangled four-photon cluster state is generated~\cite{PhysRevLett.100.210501,Zhang:16} as follows
\begin{equation}
\begin{aligned}
|C_4\rangle= & \frac{1}{2}(|HHHH\rangle + |HHVV \rangle +|VVHH\rangle - |VVVV\rangle) 
\end{aligned}
\end{equation}

The self-testing bound can be obtained by studying the relation between the observed violation $\beta$ of the corresponding Bell inequalities and the fidelity of target state. In Ref.~\cite{Baccari_2020}, they used a numerical procedure~\cite{PhysRevLett.117.070402} to estimate the self-testing bounds for the inequalities corresponding to the GHZ states and the ring states. The numerically estimated values are shown in Table~\ref{tab:value} for $N=3,4$. We have measured the correlators in the corresponding Bell inequalities for four multipartite graph states under three different pump power, where different pump power cause to fidelity reduction because of multi-pair emission of the SPDC process, as shown in Fig.~\ref{fig:Experiment}. Experimental results of each measurement operator are in agreement with the theoretical results (red dash lines). Meanwhile, as the fidelity of the prepared states are above 0.5, these states are close to the ideal target states. Thus, robust self-testing for three- and four-qubit graph states has been demonstrated experimentally.

\section{Experimental results}

\begin{table}
\centering
\caption{The self-testing bound values of numerical estimations for $N=3,4$. $\beta^C$ and $\beta^Q$ denote the classical bound and maximum quantum violation of the corresponding Bell inequality, respectively. $\beta^B$ and $\beta^E$ represent the theoretical nontrivial bound and the experimentally measured Bell violation, respectively.}
\label{tab:value}
\begin{tabular}{p{0.1\linewidth}p{0.05\linewidth}p{0.2\linewidth}p{0.12\linewidth}p{0.15\linewidth}} \\ 
\hline\hline
Target state & $\beta^{C}$ & $\beta^{Q}$ & $\beta^B$ & $\beta^E$ \\
\hline
$|G_3\rangle$ & 4 & $4\sqrt{2}\approx 5.657$ & 4.828 & $5.441(\pm 0.049)$ \\

$|G_4\rangle$ & 6 & $6\sqrt{2}\approx 8.485$ & 7.464 & $7.904(\pm 0.013)$ \\

$|C_3\rangle$ & 4 & $4\sqrt{2}\approx 5.657$ & 4.940 & $5.320(\pm 0.054)$
\\

$|C_4\rangle$ & 5 & $1+4\sqrt{2}\approx 6.657$ & 5.828 & $6.139(\pm 0.045)$ \\
\hline\hline
\end{tabular}
\end{table}

\begin{figure*}[!hbt]
\centering
\includegraphics[width=0.90\linewidth]{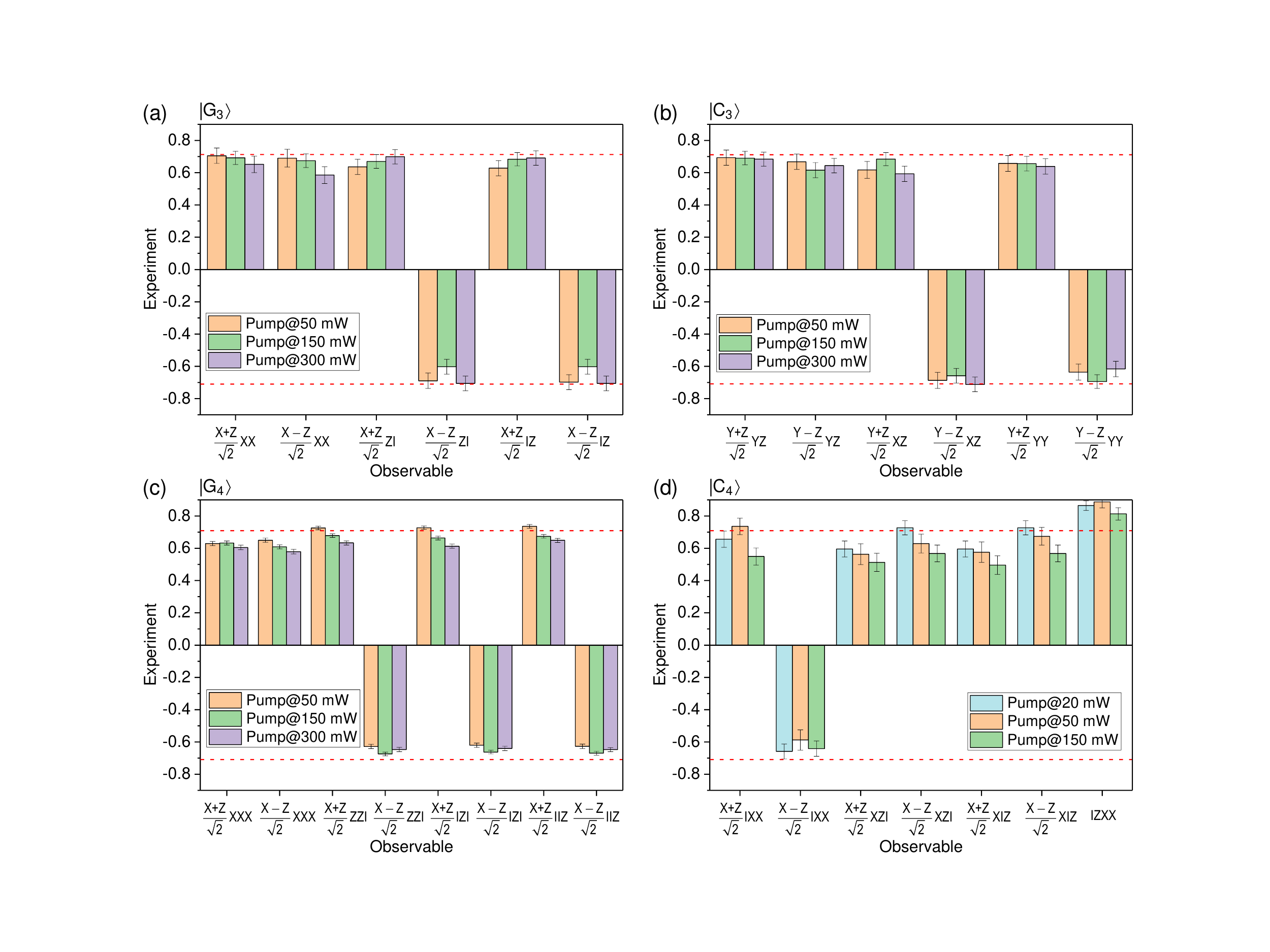}
\caption{Experimental results for self-testing graph state in different pump power. (a) and (c) The results for $|G_3\rangle=\frac{1}{\sqrt2}(|HHH\rangle+|VVV\rangle)$ and  $|G_4\rangle=\frac{1}{\sqrt2}(|HHHH\rangle+|VVVV\rangle)$. $\frac{(X\pm Z)}{\sqrt{2}}ZZI$, $\frac{(X\pm Z)}{\sqrt{2}}IZI$ and $\frac{(X\pm Z)}{\sqrt{2}}IIZ$ can be calculated from the measurement operator $\frac{(X\pm Z)}{\sqrt{2}}ZZZ$. (b) and (d) The results for linear cluster state  $|C_3\rangle=\frac{1}{\sqrt2}(|+H+\rangle+|-V-\rangle)$ and  $|C_4\rangle=\frac{1}{2}(|HHHH\rangle+|HHVV\rangle+|VVHH\rangle-|VVVV\rangle)$, where $| \pm \rangle = (|H\rangle \pm |V\rangle)/\sqrt{2}$. The value of  $\frac{(X\pm Z)}{\sqrt{2}}IXX$ can be obtained by measurement operator  $\frac{(X\pm Z)}{\sqrt{2}}ZXX$, $\frac{(X\pm Z)}{\sqrt{2}}XZI$ and $\frac{(X\pm Z)}{\sqrt{2}}XIZ$ are calculated by  $\frac{(X\pm Z)}{\sqrt{2}}XZZ$. The red dash lines represent the theoretical results ($\pm \frac{\sqrt{2}}{2}$) of these measurement operator, except for $IZXX$ (+1).}
\label{fig:Experiment}
\end{figure*}

\begin{figure}[!htp]
\centering
    \includegraphics[width=0.8\linewidth]{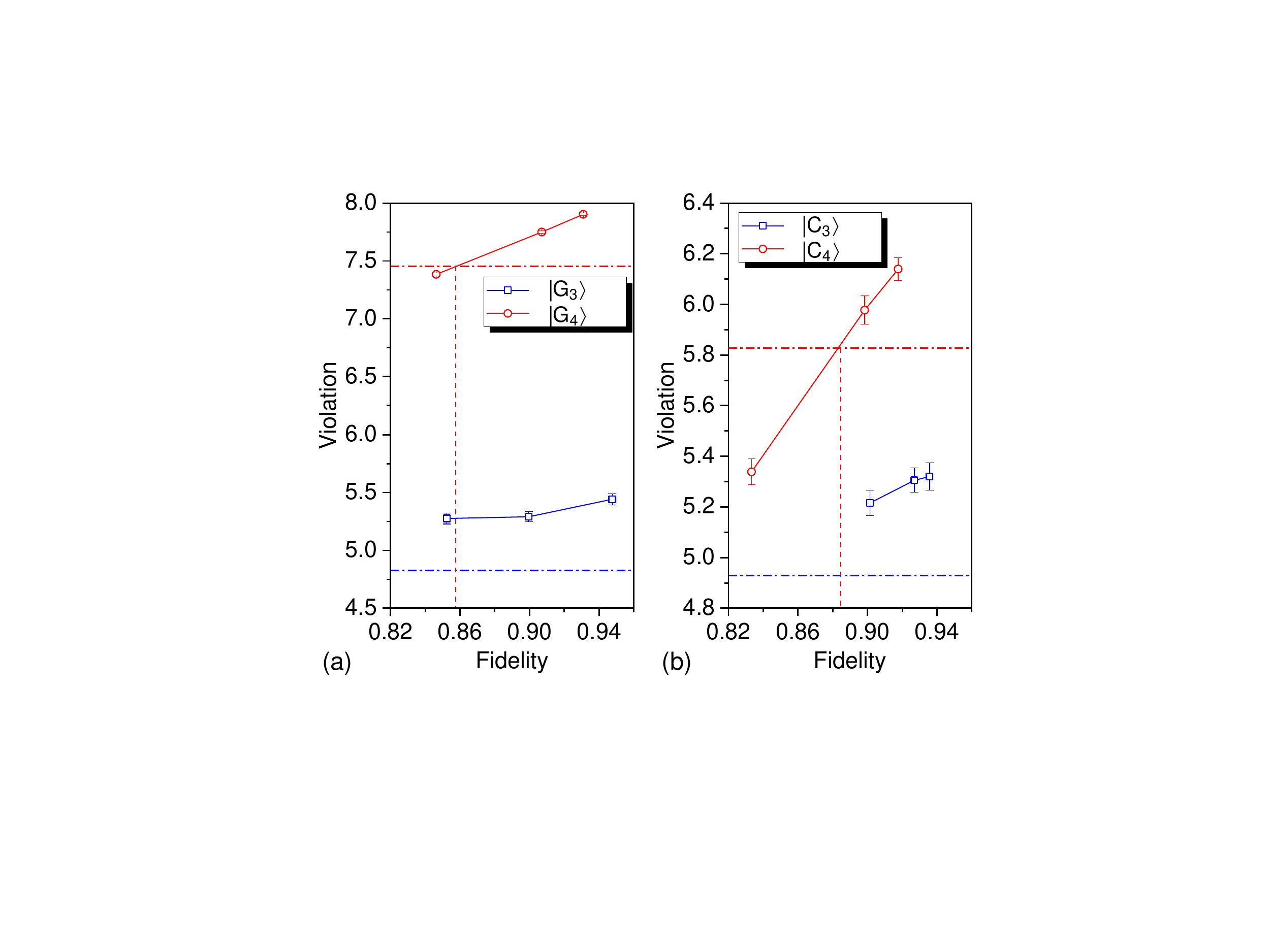}
    \caption{The relation of Bell violation to fidelity of prepared graph state. (a) For the GHZ states $|G_3\rangle$ and $|G_4\rangle$. (b) For the linear cluster states $|C_3\rangle$ and $|C_4\rangle$. Blue dash-dotted line and red one denote the values that sufficiently obtain nontrivial bounds for three-photon states and four-photon states, respectively. As shown in Table~\ref{tab:value}.
    }
    \label{fig4}
\end{figure}

From the experimental results in Fig.~\ref{fig:Experiment}, the values of Bell violation can be derived as shown in Table~\ref{tab:value}, which are greatly larger than the classical bounds $\beta^C$ and close to the maximum violations $\beta^Q$. Meanwhile, they are also noticeably higher than the nontrivial self-testing bounds $\beta^B$. Thus, the tested states are cerify to the ideal target states (GHZ and linear cluster states) under the tolerant noise. Moreover, we also measure and calculate the violations under three different noise (i.e., the fidelities of prepared target state are different), as shown in Fig.~\ref{fig4}. To calculate the distance between the prepared states and target states, one needs to obtain the fidelities of GHZ states and linear cluster states.

To obtain the fidelity of three- and four-qubit GHZ states, we express the density matrix of the GHZ states as~\cite{PhysRevLett.92.087902,PhysRevA.76.030305}
\begin{equation}
\begin{aligned}
|G_N\rangle \langle G_N| &=  \frac{1}{2}P^N+\frac{1}{2N}\sum\limits_{k=0}^{N-1}(-1)^kM_k \\
F(\rho_{exp})&=\langle G_N|\rho_{exp}|G_N\rangle
\end{aligned}
\end{equation}
where $P^N=|H\rangle \langle H|^{\otimes N}+|V\rangle \langle V|^{\otimes N}$ and $M_k=[\cos (k\pi/N) X + \sin (k\pi/N) Y]^{\otimes N}$. Expectation of $P^N$ and $M_k$ can be calculated under the computational basis $|H\rangle$, $|V\rangle$ and $(|H\rangle \pm e^{i k\pi/N} |V\rangle)/\sqrt 2$. We calculate the fidelities of three- and four-qubit GHZ states from the experimental raw data, as shown in Fig~\ref{fig4}.
For cluster state $|C_3\rangle$ and $|C_4\rangle$, we configure the fidelity $F=Tr(|C\rangle \langle C|\rho_{exp})$ by measuring their stabilizers, where $\rho_{exp}$ represent the prepared states. For cluster states, its project operator can be completely described by their stabilizers. The stabilizing operator of three-qubit cluster states reads:
\begin{equation}
\begin{aligned}
g_1&=X_1Z_2I_3 \\
g_2&=Z_1X_2Z_3 \\
g_3&=I_1Z_2X_3
\end{aligned}
\end{equation}
where $X_i,Z_i$ and $I_i$ denote Pauli operator $\sigma_x,\sigma_z$ and identity $\mathrm{I}$, these stabilizers $g_i$ and their products together form stabilizer group called $\mathrm{S}$. The target state $|C \rangle$ is defined by $g_i|C \rangle=|C\rangle(i=1,2,3)$. The projector can be written as the $2^N$ elements in the group $S$, here in the three-qubit case, $|C\rangle_3 \langle C|=\frac{1}{2^3}\sum_{\sigma\in \mathrm{S}}\sigma$. Thus, the fidelity of the target states can be obtained by calculating the average expectation value of all stabilizers.
\begin{equation}
\begin{aligned}
F_{|C_3\rangle} = & \frac{1}{8}(XZI+ZXZ+IZX+XIX+ZYY+YYZ-YXY+I)
\end{aligned}
\end{equation}

The stabilizing operator of four-qubit cluster states are:
\begin{equation}
\begin{aligned}
g_1&=X_1X_2Z_3I_4 \\
g_2&=Z_1Z_2I_3I_4 \\
g_3&=I_1Z_2X_3X_4    \\
g_3&=I_1I_2Z_3Z_4
\end{aligned}
\end{equation}
one obtains the fidelity of four-qubit cluster state by $F=Tr|C_4\rangle \langle C_4|\rho_{exp})$, the 16 expectation values can be deduced from 9 joint measurement settings.
\begin{equation}
\begin{aligned}
F_{|C_4\rangle}= & \frac{1}{16}(ZZII-YYZI+XXIZ-ZIYY +XXZI+ZIXX-IZYY+XYXY \\
&+IZXX+ZZZZ+YXYX+YXXY+IIZZ+XYYX-YYIZ+I)
\end{aligned}
\end{equation}

As long as the prepared graph states have higher fidelity, the larger Bell violation can be observed. Taking $|C_4\rangle$ as an example, the violation $\beta^E$ is beyond nontrivial bound $\beta^B=5.828$ when the fidelities of prepared graph states are $0.909\pm0.008$ and $0.919\pm 0.006$. However, the violation cannot reach the nontrivial bound for a fidelity of $0.828\pm 0.009$, which is above 0.5, implying that the corresponding prepared state is still certify to $|C_4\rangle$ under the real experimental noise.

\section{DISCUSSION AND CONCLUSION}
Multipartite graph states as an essential quantum resource, not only can be used in measurement-based quantum computation, but also play a vital role in large-scale quantum network. To guarantee its quality, an effective certification method is necessary. Self-testing as a device-independent verification tool is extremely suitable for this task. It can reveal the structure of unknown quantum state, based on the observable correlations. Especially, a robust version of self-testing is significant from the experimental point of view. Here we experimentally verified the robust self-testing protocols for multipartite graph states based on the scalable Bell inequalities. Our work paves a way toward the certification of many-body quantum system device-independently. Furthermore, high-fidelity cluster states can be tested under the minimum measurement, which substantially reduce the sources for one-way quantum computation.

\textit{Note added}.-Recently, we became aware of an independent experiment~\cite{wu2021robust}.

\section*{Funding}
National Key R\&D Program of China (2019YFA0308700, 2020YFA0309500); National Natural Science Foundation of China (11922406); Key R\&D Program of Guangdong Province (2020B0303010001).

\section*{Acknowledgments}
The authors thank A. Acín and F. Baccari for valuable discussions.

\section*{Disclosures}

The authors declare no conflicts of interest.





\end{document}